\documentclass[10pt,english,notitlepage]{revtex4-1}
\usepackage[T1]{fontenc}
\usepackage[latin9]{inputenc}
\usepackage{bm}
\usepackage{amsmath}
\usepackage{amssymb}
\usepackage{esint}

\makeatletter
 
 \@ifundefined{textcolor}{}
 {%
   \definecolor{BLACK}{gray}{0}
   \definecolor{WHITE}{gray}{1}
   \definecolor{RED}{rgb}{1,0,0}
   \definecolor{GREEN}{rgb}{0,1,0}
   \definecolor{BLUE}{rgb}{0,0,1}
   \definecolor{CYAN}{cmyk}{1,0,0,0}
   \definecolor{MAGENTA}{cmyk}{0,1,0,0}
   \definecolor{YELLOW}{cmyk}{0,0,1,0}
 }

\def\Mpl{M_{\rm Pl}}

\usepackage{babel}

\makeatother

\begin{document}

\title{Reheating in 3-form inflation}

\author{Antonio De Felice}
\affiliation{ThEP's CRL, NEP, The Institute for Fundamental Study,
Naresuan University, Phitsanulok 65000, Thailand}
\affiliation{Thailand Center of Excellence in Physics, Ministry of Education,
Bangkok 10400, Thailand}

\author{Khamphee Karwan}
\affiliation{ThEP's CRL, NEP, The Institute for Fundamental Study,
Naresuan University, Phitsanulok 65000, Thailand}
\affiliation{Thailand Center of Excellence in Physics, Ministry of Education,
Bangkok 10400, Thailand}

\author{Pitayuth Wongjun}
\affiliation{ThEP's CRL, NEP, The Institute for Fundamental Study,
Naresuan University, Phitsanulok 65000, Thailand}
\affiliation{Thailand Center of Excellence in Physics, Ministry of Education,
Bangkok 10400, Thailand}

\date{\today}
\begin{abstract}
 We consider the 3-form field, which has been considered as a candidate for realizing inflation, coupled to a scalar field which models the relativistic matter particles produced during the reheating epoch. We have investigated the stability conditions for this theory and found that introducing such a coupling does not lead to any ghosts or Laplacian instabilities. We have also investigated the reheating temperature and the production of particles due to parametric resonances. We have found that this process is more efficient in this theory compared to the result of the standard-scalar-field inflationary scenario.
\end{abstract}
\maketitle

\section{Introduction}

The study of cosmology in the last few years has become more and more
interesting and fundamental, because available data have
started giving us a non-trivial picture of the universe at large
scales. For example, data led to the unexpected result of the acceleration of the universe \cite{Riess,Constitution,WMAP1,WMAP7,BAO1,Percival,Suzuki}. In the coming years, we may expect new surprises coming
also in the field of inflation and inflationary non-gaussianities \cite{Ade:2011ah}.

In the context of early cosmology, the theoretical predictions have strong connections with fundamental
physics. However, we still do not have a clear picture of high-energy physics at energies above electroweak scale. Therefore, in inflation, many different models have been introduced. People have recently tried to search for alternative models of inflation which, on one hand, would still give account for the standard inflationary results, and, on the other hand, would leave peculiar and unique-to-the-model imprints in the data.

Since data are getting more and more precise, some of these models have been already excluded, whereas others are still viable. In fact, different models would give different values for each inflationary observable, so that in principle, as more data become available, the parameter space in model-space will consistently reduce.

In the light of high-energy physics and in order to explore the parameter space of models which are not built only of fundamental scalar fields, the 3-form inflationary field has been introduced \cite{Germani:2009iq, Koivisto:2009sd}. Dynamical models for the 3-form in cosmological backgrounds have been studied \cite{Germani:2009gg, Koivisto:2009fb, Koivisto:2009ew, Ngampitipan:2011se, DeFelice:2012jt,nong3F, Koivisto:2011rm, Urban:2012ib}. In particular, the background evolution during inflation was found to be similar to the one of the standard scalar-field case. However, at the level of perturbations, things change considerably \cite{Koivisto:2009fb, DeFelice:2012jt}. First, besides the gravitational
wave modes---which have standard features---only one scalar field
does propagate. Second, the potential for the 3-form must be chosen
such that this scalar mode degree of freedom does not become a ghost.
Third, the speed of propagation of such a scalar field, is not equal
to one, and for a class of potentials it may become even negative (leading to a Laplacian instability). In
order to prevent this latter possibility to happen, further conditions on the
potential must be chosen. Finally, because of this non-trivial speed
of propagation, these models can---depending on the form of the potential---lead
to large values of non-gaussianities, giving rise to a whole new phenomenological
picture for this 3-form inflation theory\cite{DeFelice:2012jt,nong3F}.

In the standard picture of the early universe, after inflation, the universe passes through the reheating era. It is assumed that, at that time, all elementary particles are created and then the universe enters the high temperature phase. In order to generate particles during the reheating time, the inflaton field must couple to some matter field and, consequently, decay to generate particles. This mechanism has been widely studied in the context of modeling the matter fields via a scalar field \cite{Dolgov:1982th,Abbott:1982hn,Albrecht:1982mp, Kofman:1994rk,Shtanov:1994ce,Kofman:1997yn} (for more recent reviews see e.g.\ \cite{Martin:2004um, Bassett:2005xm, Allahverdi:2010xz}) and $f(R)$ theory \cite{Mijic:1986iv, Appleby:2009uf, DeFelice:2010aj, Motohashi:2012tt}.

In this paper we investigate the role that the 3-form field can play in order to reheat the universe. The model, its definition, and the coupling with matter fields are described in Section \ref{sec:The-Lagrangian}.
Then, in Section \ref{sec:Perturbation-theory}, we study the conditions for the theory to be stable against ghosts (i.e.\ fields with a negative kinetic energy) and Laplacian instabilities (i.e.~$c_{s}^{2}<0$) when such a 3-form is coupled with a matter field. Section \ref{sec:Reheating-phase} is devoted to the study
of the background approximate solution during the reheating phase. We finally find an estimated expression for the reheating temperature and discuss the preheating phase in comparison to other already known and studied models.

\section{The Lagrangian\label{sec:The-Lagrangian}}

We will study the Lagrangian for a 3-form field, $A_{\mu \nu \rho}$, \cite{Koivisto:2009fb, Koivisto:2009ew, Ngampitipan:2011se, DeFelice:2012jt}  coupled to a scalar field, $\phi$, given as follows
\begin{eqnarray}
S & = & \int d^{4}x\sqrt{-g}\left[\frac{\Mpl^{2}}{2}\, R-\frac{1}{48}\, F_{\alpha\beta\gamma\delta}F^{\alpha\beta\gamma\delta}-V(A_{\alpha\beta\gamma}A^{\alpha\beta\gamma})\right.\nonumber \\
 &  & \left.{}-\frac{1}{2}\nabla_{\mu}\phi\nabla^{\mu}\phi-\frac{1}{2}\, m_{\phi}^{2}\phi^{2}-\frac{\lambda}{6}\, E^{\mu\alpha\beta\gamma}A_{\alpha\beta\gamma}\phi\nabla_{\mu}\phi\right],\label{action}
\end{eqnarray}
where $E_{\alpha\beta\gamma\delta}$ is the Levi-Civita antisymmetric tensor on curved backgrounds, which on Minkowski reduces to $\epsilon_{\alpha\beta\gamma\delta}$ (with $\epsilon_{0123}=1=-\epsilon^{0123}$). Then we also have $E^{\alpha\beta\gamma\delta}=\epsilon^{\alpha\beta\gamma\delta}/\sqrt{-g}$. The last term is a coupling term between the 3-form and the scalar field (which mimics a relativistic matter field) into which the 3-form
decays. This coupling term is one of the simplest one we may think of, and it is the lowest dimensional analytical one (in fact here $\lambda$ is a dimension-less parameter), which can introduce a decay of the 3-form field into two scalar field particles. Other possible coupling terms may arise but we expect their coupling constants to be suppressed by the cutoff scale of the theory. By integration by parts, we have that the considered coupling term can be rewritten as
\begin{equation}
\frac{\lambda}{2}\,\phi^{2}\,\nabla_{\mu}B^{\mu}\,,
\end{equation}
where $B^{\mu}$ is the vector dual to the 3-form, that is $B^{\mu}=E^{\mu\alpha\beta\gamma}A_{\alpha\beta\gamma}/3!$.
The coupling term can be rewritten as
\begin{equation}
-\frac{\lambda}{6}\, E^{\mu\alpha\beta\gamma}A_{\alpha\beta\gamma}\phi\nabla_{\mu}\phi=-\frac{\lambda\phi}{\sqrt{-g}}\left[-A_{123}\partial_{0}\phi+A_{023}\partial_{1}\phi-A_{013}\partial_{2}\phi+A_{012}\partial_{3}\phi\right],
\end{equation}
which explicitly shows the coupling between the scalar field and the
four independent components of the 3-form.

 We assume a flat Friedmann-Lema\^\i tre-Robertson-Walker (FLRW) manifold
\begin{equation}
ds^{2}=-dt^{2}+a(t)^{2}d\bm{x}^{2}\,,
\end{equation}
and, on this background, we will set the
the components of the 3-form $A_{\alpha\beta\gamma}$ compatibly with the background symmetries and with the field equations of motion, as in
\begin{equation}
A_{0ij}=0\,,\qquad A_{ijk}=a^{3}\epsilon_{ijk}\, X\,,
\end{equation}
where $\epsilon_{ijk}$ is the three-dimensional Levi-Civita symbol
(with $\epsilon_{123}=1$). The Friedmann equation of motion reads as
\begin{equation}
E_{1}\equiv3\Mpl^{2}H^{2}-\rho_{X}-\rho_{\phi}=0\,,\label{eq:fried}
\end{equation}
where $H\equiv \dot{a}/a$ is the Hubble parameter, and
\begin{eqnarray}
\rho_{X} & = & \frac{1}{2}\,\dot{X}^{2}+V+\frac{9}{2}\, H^{2}X^{2}+3HX\dot{X}=\frac{1}{2}\, Y^{2}+V\,,\label{rhoX}\\
\rho_{\phi} & = & \frac{1}{2}\,\dot{\phi}^{2}+\frac{1}{2}\, m_{\phi}^{2}\phi^{2}\,,\\
Y & = & \dot{X}+3HX\,.\label{Y-X}
\end{eqnarray}
We also have the second Einstein equation as
\begin{equation}
E_{2}\equiv\Mpl^{2}(2\dot{H}+3H^{2})+p_{X}+p_{\phi}=0\,,\label{eq:acc}
\end{equation}
where $p_{X}$ is the 3-form effective pressure defined as follows
\begin{eqnarray}
p_{X} & = & -\left(\frac{1}{2}\,\dot{X}^{2}+V+3HX\dot{X}+\frac{9}{2}\, H^{2}X^{2}-12V_{,y}X^{2}\right)\,.\\
p_{\phi} & = & \frac{1}{2}\,\dot{\phi}^{2}-\frac{1}{2}\, m_{\phi}^{2}\phi^{2}\,.
\end{eqnarray}
The equation of motion for $X$ can be written as
\begin{equation}
E_{X}\equiv\ddot{X}+3(H\dot{X}+X\dot{H})+12V_{,y}X-\lambda\phi\dot{\phi}=0\,.\label{X Eq}
\end{equation}
Finally, the equation of motion for the scalar matter can be written
as
\begin{equation}
E_{\phi}\equiv\ddot{\phi}+3H\dot{\phi}+m_{\phi}^{2}\phi+\lambda\phi Y=0\,.
\end{equation}
The Bianchi identities lead to
\begin{equation}
\dot{E}_{1}+3H(E_{1}-E_{2})+YE_{X}+\dot{\phi}E_{\phi}=0\,.
\end{equation}

\section{Perturbation theory\label{sec:Perturbation-theory}}

In order to study the no-ghost conditions, from the action approach, it is convenient, though not necessary, to choose a gauge for which the spatial metric is diagonal (i.e.~flat
gauge) \cite{cosmoper}, so that the metric tensor can be written as
\begin{equation}
ds^{2}=-(1+2\alpha)dt^{2}+2\partial_{i}\psi\, dt\, dx^{i}+a^{2}(1+2\Phi)\, d\bm{x}^{2}\,.
\end{equation}
As for the 3-form, we can use
a time gauge to fix its scalar perturbations as \cite{Koivisto:2009fb, DeFelice:2012jt}
\begin{equation}
A_{0ij}=a\epsilon_{ijk}\partial_{k}\beta(t,\bm{x})\,,\qquad A_{ijk}=a^{3}\epsilon_{ijk}\, X(t)\,.
\end{equation}
Finally we will perturb the matter scalar field as $\phi=\phi(t)+\delta\phi$.
By expanding the action at second order we find

\begin{eqnarray}
S & = & \int dt\, a^{3}\left\{ \frac{(\partial^{2}\beta)^{2}}{2a^{4}}+6V_{,y}\frac{(\partial\beta)^{2}}{a^{2}}+\left[\lambda\phi\delta\phi+12V_{,y}X\psi+Y(\alpha+3\Phi)\right]\frac{\partial^{2}\beta}{a^{2}}\right.\nonumber \\
 &  & {}+6V_{,y}X^{2}\frac{(\partial\psi)^{2}}{a^{2}}-\left[2\Mpl^{2}(H\alpha-\dot{\Phi})-\dot{\phi}\delta\phi\right]\frac{\partial^{2}\psi}{a^{2}}\nonumber \\
 &  & {}-\frac{1}{2}\left(6\Mpl^{2}H^{2}-Y^{2}-\dot{\phi}^{2}\right)\alpha^{2}-3\Mpl^{2}\dot{\Phi}^{2}+\frac{1}{2}\,\dot{\delta\phi}^{2}\nonumber \\
 &  & {}+\left[6\Mpl^{2}H\dot{\Phi}-2\Mpl^{2}\frac{\partial^{2}\Phi}{a^{2}}+3(Y^{2}+12V_{,y}X^{2})\Phi-\dot{\phi}\dot{\delta\phi}-m_{\phi}^{2}\phi\,\delta\phi\right]\alpha\nonumber \\
 &  & {}+\Mpl^{2}\frac{(\partial\Phi)^{2}}{a^{2}}-\frac{9}{2}\left(12V_{,y}X^{2}-Y^{2}+144V_{,yy}X^{4}\right)\Phi^{2}\nonumber \\
 &  & {}+\left.3\left[\dot{\phi}\dot{\delta\phi}-m_{\phi}^{2}\phi\delta\phi\right]\Phi-\frac{1}{2}\left(\lambda Y+m_{\phi}^{2}\right)\delta\phi^{2}-\frac{1}{2}\frac{(\partial\delta\phi)^{2}}{a^{2}}\right\} ,
\end{eqnarray}
where we have defined $y=A_{\alpha\beta\gamma} A^{\alpha\beta\gamma}=6X^2$, on the background. By integrating out the auxiliary fields one finds two no-ghost conditions
for the remaining two propagating fields ($\Phi$ and $\delta\phi$).
The independent no-ghost conditions for the kinetic matrix can then be
written as
\begin{eqnarray}
A_{22} & = & \frac{1}{2}\,{\frac{\Mpl^{2}{a}^{3}H\,[6\,{a}^{2}\left(2\, H\Mpl^{2}+3\,{X}^{2}H-2\, XY\right)V_{{,y}}+\Mpl^{2}H{k}^{2}]}{\Mpl^{4}{H}^{2}{k}^{2}-3\,{a}^{2}\left({X}^{2}{\it \dot{\phi}}^{2}-4\,{H}^{2}\Mpl^{4}+4\,\Mpl^{2}XHY-6\,\Mpl^{2}{X}^{2}{H}^{2}\right)V_{{,y}}}}\,,\\
\det A & = & A_{11}A_{22}-A_{12}^{2}={\frac{3\Mpl^{4}{a}^{8}V_{{,y}}{Y}^{2}}{\Mpl^{4}{H}^{2}{k}^{2}-3\,{a}^{2}\left({X}^{2}{\it \dot{\phi}}^{2}-4\,{H}^{2}\Mpl^{4}+4\,\Mpl^{2}XHY-6\,\Mpl^{2}{X}^{2}{H}^{2}\right)V_{{,y}}}}\,.
\end{eqnarray}
Since
\begin{equation}
4\,{H}^{2}\Mpl^{4}-4\,\Mpl^{2}XHY+6\,\Mpl^{2}{X}^{2}{H}^{2}=\frac{4\Mpl^{2}}{3}\,\bigl(\tfrac{1}{2}\,\dot{X}^{2}+V+\rho_{\phi}\bigr)\,,
\end{equation}
we can change the denominator of the previous two expressions as
\begin{eqnarray}
\det A & = & {\frac{3\Mpl^{4}{a}^{8}V_{{,y}}{Y}^{2}}{\Mpl^{4}{H}^{2}{k}^{2}+{a}^{2}\, V_{{,y}}\,\bigl[4\Mpl^{2}\bigl(\frac{1}{2}\dot{X}^{2}+V+\rho_{\phi}\bigr)-3{X}^{2}{\it \dot{\phi}}^{2}\bigr]}}>0\,,\label{eq:ghost1}\\
A_{22} & = & \frac{1}{2}\,{\frac{{a}^{3}\,[4\Mpl^{2}{a}^{2}\,\bigl(\tfrac{1}{2}\,\dot{X}^{2}+V+\rho_{\phi}\bigr)V_{{,y}}+\Mpl^{4}H^{2}{k}^{2}]}{\Mpl^{4}{H}^{2}{k}^{2}+{a}^{2}\, V_{{,y}}\,\bigl[4\Mpl^{2}\bigl(\frac{1}{2}\dot{X}^{2}+V+\rho_{\phi}\bigr)-3{X}^{2}{\it \dot{\phi}}^{2}\bigr]}}>0\,.\label{eq:ghost2}
\end{eqnarray}
These results show that the coupling does not directly contribute
to the ghost condition. Condition (\ref{eq:ghost1}), for high $k$'s,
is satisfied when
\begin{equation}
V_{,y}>0\,,
\end{equation}
which corresponds to the no-ghost condition already found in the vacuum
case (i.e.~in the absence of the $\phi$ field) \cite{DeFelice:2012jt}. Furthermore, we
now require also
\begin{equation}
4\Mpl^{2}\bigl(\tfrac{1}{2}\dot{X}^{2}+V+\rho_{\phi}\bigr)-3{X}^{2}{\it \dot{\phi}}^{2}=2\Mpl^{2}\,(\dot{X}^{2}+2V)+(2\Mpl^{2}-3{X}^{2}){\it \dot{\phi}}^{2}+2\Mpl^{2}m_{\phi}^{2}\phi^{2}>0\,.\label{eq:ghlowk}
\end{equation}
Sufficient condition for Eq.~(\ref{eq:ghlowk}) to be verified is
imposing $V\geq0$ (the other condition found in the vacuum case \cite{DeFelice:2012jt}),
and $|X/\Mpl|\leq\sqrt{2/3}$. During reheating (i.e.\ after inflation ends) this condition is
satisfied as, in general, $X/\Mpl\to0$.

In order to check whether this condition is violated or not also during
inflation, we change the variables to the dimensionless variables
as
\begin{equation}
x=\frac{X}{\Mpl},\,\,\,\, w^{2}=\frac{Y^{2}}{6H^{2}\Mpl^{2}},\,\,\,\, z^{2}=\frac{V}{3H^{2}\Mpl^{2}},\,\,\,\, u^{2}=\frac{\dot{\phi}^{2}}{6H^{2}\Mpl^{2}},\,\,\,\, v^{2}=\frac{m_{\phi}^{2}\phi^{2}}{6H^{2}\Mpl^{2}}.
\end{equation}
Therefore, the no-ghost condition becomes
\begin{equation}
(\sqrt{2}-\sqrt{3}x)^{2}+(2-3x^{2})u^{2}+2z^{2}+2v^{2}>0\,,
\end{equation}
where we have used $w^{2}\sim 1$ which is the requirement for inflation
\cite{DeFelice:2012jt}.
According to the Friedmann equation (\ref{eq:fried}), this requirement
implies that $z^{2}\ll1$ when the inflation is supposed to be driven
only by the 3-form field. Moreover, since the contribution of the
scalar field $\phi$ to the dynamics of inflationary universe is sub-dominant,
the values of $u^{2}$ and $v^{2}$ need to be smaller than $z^{2}$.
By taking into account both the first and the second terms, it is clear
that the no-ghost condition is satisfied when $x\leq\sqrt{2/3}$
and $x\gg\sqrt{2/3}$. However, for $x\gtrsim\sqrt{2/3}$ these terms
will take a negative value. In order to check that this negative value
violates the no-ghost condition or not, one can find the minimum of
the function $f(x)$, by keeping $u$ constant, as in $f(x)=(\sqrt{2}-\sqrt{3}x)^{2}+(2-3x^{2})u^{2}$.
This minimum value takes the form $f_{\text{min}}=-2u^{4}/(1-u^{2})\approx-2u^{4}.$
Now it is clear that the no-ghost condition is satisfied for all ranges
of $x$ since $f_{\text{min}}+2z^{2}+2v^{2}>0$, as $u^{2}\ll z^{2}\ll1$.

\subsection{Speeds of propagation}

In order to find the two speeds of propagation, we can proceed as
follows. We perform the following field redefinition
\begin{eqnarray}
\Phi & = & k\, q_{1}\,,\\
\delta\phi & = & -\frac{A_{12}}{A_{22}}\, k\, q_{1}+q_{2}={\frac{\dot{\phi}\left[6\,{a}^{2}\left(2\,\Mpl^{2}H+3\, H{X}^{2}-XY\right)V_{{,y}}+\Mpl^{2}H{k}^{2}\right]}{H\left[6\,{a}^{2}\left(2\,\Mpl^{2}H+3\, H{X}^{2}-2\, XY\right)V_{{,y}}+\Mpl^{2}H{k}^{2}\right]}}\, k\, q_{1}+q_{2}\,.
\end{eqnarray}
In this case, the second order action, after a few integrations by
parts, reduces to
\begin{equation}
S=\int dt\,[Q_{ij}\dot{q}_{i}\dot{q_{j}}-C_{ij}q_{i}q_{j}-B_{ij}q_{i}\dot{q}_{j}]\,,
\end{equation}
where the matrix $Q_{ij}$ is diagonal (without approximations, due
to the proposed field redefinition) and its diagonal elements, for
large $k$'s are only background dependent as they reduce to
\begin{equation}
Q_{11}\simeq\frac{6a^{5}V_{,y}Y^{2}}{H^{2}}\,,\qquad Q_{12}=0\,,\qquad Q_{22}=\frac{a^{3}}{2}\,.
\end{equation}
Along the same lines, by using the equations of motion, we can prove
that at order $k^{2}$, the matrix elements $C_{ij}$ reduce to
\begin{equation}
C_{11}\simeq\frac{6a^{3}Y^{2}(V_{,y}+12V_{,yy}X^{2})\, k^{2}}{H^{2}}\,,\qquad C_{12}\simeq0\,,\qquad C_{22}\simeq\frac{ak^{2}}{2}\,.
\end{equation}
The antisymmetric matrix $B_{ij}$, by using the equations of motion,
is of order ${\cal O}(k^{-1})$ and can be neglected, as long as we
consider only the high-$k$ behavior of the theory.

Therefore the two speeds of propagation are
\begin{equation}
c_{X}^{2}=1+\frac{12V_{,yy}X^{2}}{V_{,y}}\,,\qquad c_{\phi}^{2}=1\,.
\end{equation}
This result is consistent with the result in vacuum \cite{DeFelice:2012jt} and with what we would naively expect, due to the form of the coupling we have chosen. Since after the end of inflation the ghost-free potentials studied in \cite{DeFelice:2012jt} all reduce to $V\approx\tfrac12 m^2 X^2=\tfrac1{12}m^2y$, then during reheating there are no Laplacian instabilities as $c_X^2\to1$.

\section{Reheating phase\label{sec:Reheating-phase}}

Let us discuss about the behavior of the 3-form field during the reheating
phase. For simplicity, we will neglect the backreaction effects of the scalar-matter
field on the leading term in the background expansion. Because of
this approximation, we can neglect the interaction term in Eq.~(\ref{X Eq})
to find a solution of this equation. By differentiating Eq.~(\ref{X Eq}),
one can write this equation in terms of $Y$ as
\begin{equation}
\ddot{Y}+3H\dot{Y}+m^{2}Y=0.\label{Y Eq}
\end{equation}
Here, we have used the potential form $V=\frac{1}{2}m^{2}X^{2}$ for
simplicity. This potential form is free from any ghost and Laplacian
instabilities during inflation \cite{DeFelice:2012jt}, since the field
$X$ is always less than $\sqrt{2/3}\Mpl$. Moreover, when $X/\Mpl<1$, as already stated above,
all potentials which are free from ghosts and Laplacian instabilities
at the end of inflation approximately take this quadratic form.

From Eq.~(\ref{Y Eq}), it follows that $Y^{2}\propto H^{2}\gg|\dot{H}|$
when $m^{2}\ll H^{2}$, so that the three-form field can indeed drive
inflation. During inflation, $H$ slowly decreases since $\dot{H}<0$,
fulfilling in this way the no-ghost condition, up to the end of inflation,
at scales around $m^{2}\sim H^{2}$. After the end of inflation, the
reheating phase starts as both the fields $X$ and $Y$ begin to oscillate
at scales $m^{2}\gg H^{2}$. The evolution of $Y$ and $X$ during
the oscillating phase can be studied by changing variable $Y=a^{-3/2}\bar{Y}$,
so that Eq.~(\ref{Y Eq}) becomes
\begin{equation}
\ddot{\bar{Y}}+\left(m^{2}-\frac{9}{4}H^{2}-\frac{3}{2}\dot{H}\right)\bar{Y}=0.\label{Ybar Eq}
\end{equation}
Using the approximations $m^{2}\gg H^{2}$ and $m^{2}\gg|\dot{H}|$,
the solution for Eq.~(\ref{Ybar Eq}) can be written as
\begin{equation}
\bar{Y}=C\sin(mt),
\end{equation}
where $C$ is a constant. Thus, the solution of $Y$ can be written
as
\begin{equation}
Y=Ca^{-3/2}\sin(mt).\label{Y_sol_a}
\end{equation}
By neglecting the contribution from $\phi$ in Eq.~(\ref{eq:fried})
and Eq.~(\ref{eq:acc}) and combining them together, one obtains
\begin{equation}
2\Mpl^{2}\left[\frac{\ddot{a}}{a}+2\left(\frac{\dot{a}}{a}\right)^{2}\right]=Y^{2}=C^{2}a^{-3}\sin^{2}(mt).\label{eq_for_a}
\end{equation}
The solution for this equation is
\begin{equation}
a(t)=\left[c_{2}t+3c_{1}+\frac{3C^{2}[2m^{2}t^{2}+\cos(2mt)]}{16m^{2}\Mpl^{2}}\right]^{1/3},
\end{equation}
where $c_{1}$ and $c_{2}$ are integration constants. During the
oscillating phase, $mt\gg1$, so that the scale factor can be approximated
as
\begin{equation}
a(t)\simeq\left(\frac{3C^{2}t^{2}}{8\Mpl^{2}}\right)^{1/3}.\label{a sol}
\end{equation}
By substituting this approximated solution into Eq.~(\ref{Y_sol_a}),
one obtains
\begin{equation}
Y=\sqrt{\frac{8}{3}}\Mpl\,\frac{\sin(mt)}{t}\,.\label{Ysolution}
\end{equation}
On using the solution for $a$ in Eq.~(\ref{a sol}), the expression
for $H$ becomes
\begin{equation}
H=\frac{2m\{8c_{2}m\Mpl^{2}-3C^{2}[\sin(2mt)-2mt]\}}{48m^{2}(c_{2}t+3c_{1})\Mpl^{2}+9C^{2}[2m^{2}t^{2}+\cos(2mt)]}\,.
\end{equation}
Therefore, the approximated solution during oscillating phase, $mt\gg1$,
reads
\begin{equation}
H\simeq\frac{2}{3t}\,.
\end{equation}
Note that this estimated value for $H$ agrees with the numerical
result in \cite{DeFelice:2012jt} since $H^{2}\propto a^{-3}=t^{-2}$.
Substituting $H$ into Eq.~(\ref{X Eq}) and on using the relation
in Eq.~(\ref{Y-X}), the solution for $X$ can be written as
\begin{equation}
X=\sqrt{\frac{8}{3}}\frac{\Mpl}{(mt)^{2}}\,[\sin(mt)-mt\cos(mt)]\,.
\end{equation}
Substituting $X,Y$ and $H$ into Eq.~(\ref{rhoX}), one finds that, at a first approximation,
$\rho_{X}\propto a^{-3}$. This implies, as expected, that during
oscillating phase, the 3-form field behaves as dust. This behavior
of the field also agrees with the result found in vacuum \cite{Koivisto:2009ew}.

\subsection{Quantum production of particles}

After knowing how the background evolves for the fields in the theory, we now want to find an (approximate)  expression for the value of the reheating temperature.
During the inflationary period, all matter fields are diluted away due to the
exponentially accelerating expansion of the universe. At the beginning
of the oscillating phase, all matter fields can then be assumed to
start off in their vacuum state. Hence, it is convenient to consider
the interaction between the classical background field, $X$, which
is driving inflation and the quantum scalar field $\phi$ with the
Lagrangian in the action (\ref{action}). The quantum scalar field
$\phi$ can be decomposed, due to the symmetries of the FLRW manifold,
in the Heisenberg picture, as
\begin{equation}
\hat{\phi}(t,\bm{x})=\frac{1}{(2\pi)^{3/2}}\int d^{3}k\left(\hat{a}_{k}\phi_{k}(t)e^{-i\bm{k}\cdot\bm{x}}+\hat{a}_{k}^{\dagger}\phi_{k}^{*}(t)e^{i\bm{k}\cdot\bm{x}}\right),
\end{equation}
where $\hat{a}_{k}$ and $\hat{a}_{k}^{\dagger}$ are the annihilation
and creation operators respectively, and $\bm{k}$ represents the
three dimensional wave vector. The Fourier mode $\phi_{k}(t)$ obeys
the classical equation of motion
\begin{equation}
\ddot{\phi}_{k}+3H\dot{\phi}_{k}+\left(\frac{k^{2}}{a^{2}}+m_{\phi}^{2}+\lambda Y\right)\phi_{k}=0\,.\label{eq:phik}
\end{equation}
By comparing this equation of motion to the standard scalar inflaton
field $\varphi$, with the four-legs coupling interaction $g^{2}\varphi^{2}\phi^{2}$
with matter fields, and the three-legs interaction $2g^{2}\sigma\varphi\phi^{2}$
(which arises when the field acquires a vacuum expectation value and
consequently symmetry gets broken), the modification consists of replacing
$\lambda Y$ by $g^{2}\varphi^{2}$ and $2g^{2}\sigma\varphi$ respectively,
where $\varphi=(2\Mpl/\sqrt{3})\,\sin(m_{\varphi}t)/(m_{\varphi}t)$
\cite{Dolgov:1982th,Abbott:1982hn,Albrecht:1982mp,Kofman:1994rk,Shtanov:1994ce,Kofman:1997yn}.
 On the other hand, as for the $f(R)$ models and Starobinsky inflation, defined by $f(R)=R+R^{2}/(6m^{2})$
\cite{Starobinsky:1980te}, the modification in Eq.~(\ref{eq:phik})
consists of replacing $\lambda Y$ by $\xi R$ where $R=-4m\sin(mt)/t$
\cite{Mijic:1986iv,DeFelice:2010aj}. It should be pointed out that,
considering $X$ as a classical field, implies that the coupling term
effectively contributes to the (time-dependent) mass term of the modes $\phi_{k}$
as in $\delta m^{2}=\lambda Y$.

\subsection{Reheating temperature}

In order to find an expression for the reheating temperature in our model,
we will use a strategy analogous to the one used in $f(R)$ gravity
\cite{Mijic:1986iv,DeFelice:2010aj}. First, we introduce a new variable
$u_{k}=a\phi_{k}$ and use the conformal time $\eta=\int a^{-1}dt$
as the time variable. The equation of motion for the modes of the produced
particle field can be rewritten as
\begin{equation}
\frac{d^{2}u_{k}}{d\eta^{2}}+\left[k^{2}+a^{2}m_{\phi}^{2}-\frac{1}{6}a^{2} R+\lambda a^{2}Y\right]u_{k}=0.\label{eq:u1}
\end{equation}
where $R=6a^{-3}d^{2}a/d\eta^{2}$ is the Ricci scalar.
 According to the coupling
we have introduced, a heavy particle $X$ at rest decays into two
particles $\phi$ in which both will have a total energy typically
much larger than their rest mass. In other words the produced particles
will be relativistic, that is $(k/a)^{2}\gg m_{\phi}^{2}$. In this
case, we can ignore the second term in the parentheses compared to
the first one. Since $R\propto H^{2}\sim\dot{H}$, the third
term in the parentheses is also, in general, negligible when it is
compared to the fourth one. In fact, this approximation is valid for
large $t$, since $H^{2}\propto t^{-2}$ and $Y\propto \Mpl t^{-1}$. Therefore
Eq.~(\ref{eq:u1}) will be written as
\begin{equation}
\frac{d^{2}u_{k}}{d\eta^{2}}+k^{2}u_{k}=Uu_{k},\label{eq:u2}
\end{equation}
where $U=-\lambda a^{2}Y$. For the mode deep inside the Hubble radius,
$k^{2}\gg U$, we can choose, as usual, the initial vacuum as the
state with positive-frequency solution, that is $u_{k}^{(i)}=e^{-ik\eta}/\sqrt{2k}$.
The solution of Eq.~(\ref{eq:u2}) can then be written as \cite{Zeldovich:1971mw}
\begin{equation}
u_{k}=u_{k}^{(i)}+\frac{1}{k}\int_{0}^{\eta}U(\eta')\sin[k(\eta-\eta')]u_{k}(\eta')d\eta'.\label{sol:u2}
\end{equation}
In curved spacetime, the choice of the decomposition of $\phi$ into
$\hat{a}_{k}$ and $\hat{a}_{k}^{\dagger}$ is not unique. It is possible
to use another decomposition $\hat{A}_{k}$ and $\hat{A}_{k}^{\dagger}$
which can be written in terms of $\hat{a}_{k}$ and $\hat{a}_{k}^{\dagger}$
as $\hat{A}_{k}=\alpha_{k}\hat{a}_{k}+\beta_{k}^{*}\hat{a}_{k}^{\dagger}$.
The coefficients $\alpha_{k}$ and $\beta_{k}$ coincide with the
coefficients of the Bogoliubov transformation for the ladder operators.
This transformation is chosen as to diagonalize the Hamiltonian of
the field $\hat{\phi}$ at each slicing time $\eta$. Normalization
provides the following condition for the coefficients $\alpha_{k}$
and $\beta_{k}$:
\begin{equation}
|\alpha_{k}|^{2}-|\beta_{k}|^{2}=1.
\end{equation}
If $\beta_{k}=0$, we get $\alpha_{k}=1$ and then we recover the
standard creation and annihilation operators. In general, $\beta$
will describe the produced particle due to the expansion of the universe
and can be written as \cite{Starobinsky:1981vz}
\begin{equation}
\beta_{k}=\frac{-i}{2k}\int_{0}^{\eta}U(\eta')e^{-2ik\eta'}d\eta'.\label{beta}
\end{equation}
The energy density of the produced particle in comoving coordinate
$\eta$ is defined by \cite{Zeldovich:1971mw}
\begin{eqnarray}
\rho_{\eta} & = & \frac{1}{2\pi^{2}}\int_{0}^{\infty}dkk^{2}\cdotp k|\beta_{k}|^{2},\nonumber \\
 & = & \frac{1}{8\pi^{2}}\int_{0}^{\infty}U(\eta')d\eta'\int_{0}^{\infty}U(\eta)d\eta\int_{0}^{\infty}dk~ke^{2ik(\eta'-\eta)}.\label{rho_comv}
\end{eqnarray}
By using $\int_{0}^{\infty}dk~ke^{ikx}=-1/x^{2}$ and the fact that
$U\rightarrow0$ both at early and late times, we obtain
\begin{equation}
\rho_{\eta}=\frac{1}{32\pi^{2}}\int_{0}^{\infty}\frac{dU(\eta)}{d\eta}d\eta\int_{0}^{\infty}\frac{U(\eta')}{\eta'-\eta}\, d\eta'.
\end{equation}
From $Y\propto a^{-3/2}(t)\sin(mt)\propto a^{-3/2}(\eta)\sin(\int_{0}^{\eta}ma~d\bar{\eta})$,
we can estimate $U$ as
$U=-\lambda a^{2}Y\propto-\lambda a^{1/2}(\eta)\sin(\int_{0}^{\eta}ma\, d\bar{\eta})$.
Thus we can write $U$ in terms of conformal time as
\begin{equation}
U(\eta)=Ca^{1/2}(\eta)\sin\!\left(\int_{0}^{\eta}\omega d\bar{\eta}\right),
\end{equation}
where $C$ is an overall constant factor and $\omega=ma$. Using the
approximation $a^{-1}da/d\eta\ll\omega$, valid during the rapid-oscillations
phase, we obtain
\begin{equation}
\frac{dU(\eta)}{d\eta}\cong Ca^{1/2}(\eta)\omega\cos\!\left(\int_{0}^{\eta}\omega d\bar{\eta}\right).
\end{equation}
By taking the limit $\int_{0}^{\eta}\omega d\bar{\eta}\gg1$ and using
the relation 
$\lim_{k\rightarrow\infty}\sin(kx)/x=\pi\delta(x)$, we obtain
\begin{equation}
\rho_{\eta}\cong\frac{1}{32\pi}\int_{0}^{\infty}C^{2}a\omega\cos^{2}\!\left(\int_{0}^{\eta}\omega d\bar{\eta}\right)d\eta.
\end{equation}
Note that we have cut out the infinite contribution from our integration.
Shifting the phase of the oscillating factor by $\pi/2$ and differentiating
the above equation, one obtains
\begin{equation}
\frac{d\rho_{\eta}}{dt}=\frac{1}{a}\,\frac{d\rho_{\eta}}{d\eta}\cong\frac{mU^{2}}{32\pi}=\frac{ma^{4}\lambda^{2}Y^{2}}{32\pi}.
\end{equation}
The physical energy density of the matter field $\phi$ relates to
the comoving energy density by $\rho_{\phi}=\rho_{\eta}/a^{4}$. By
taking into account the effect of the total relativistic matter degrees
of freedom, the energy density of the total radiation produced in
the reheating process takes the following form
\begin{equation}
\rho_{r}=\frac{g_{*}}{a^{4}}\rho_{\eta}=\frac{g_{*}m\lambda^{2}}{32\pi a^{4}}\int_{t_{\mathrm{os}}}^{t}a^{4}Y^{2}dt,\label{rho_r1}
\end{equation}
where $t_{\mathrm{os}}$ is a time when the oscillating phase begins.
In the regime $m(t-t_{\mathrm{os}})\gg1$, the behavior of the scale
factor and the averaged expression for $Y^{2}$ can be written as
\begin{equation}
a\simeq a_{0}(t-t_{\mathrm{os}})^{2/3},\qquad\langle Y^{2}\rangle\simeq\frac{4\Mpl^2}{3(t-t_{\mathrm{os}})^{2}}.
\end{equation}
Substituting these expressions into Eq.~(\ref{rho_r1}), we finally
obtain
\begin{equation}
\rho_{r}\simeq\frac{g_{*}\lambda^{2}}{40\pi}\,\frac{m\Mpl^2}{(t-t_{\mathrm{os}})}.\label{rho_r2}
\end{equation}
From the evolution of scale factor, one can find the evolution of
the Hubble parameter
\begin{equation}
H^{2}\simeq\frac{4}{9(t-t_{\mathrm{os}})^{2}}.\label{Hub_reh}
\end{equation}
From Eq.~(\ref{rho_r2}) and Eq.~(\ref{Hub_reh}), we  find that
the energy density $\rho_{r}$ decreases slower than $H^{2}$. Therefore
$\rho_{r}$  becomes, at some time, the dominant contribution to
the total energy density. We can estimate the duration time of the
reheating process by using the Friedmann equation $3\Mpl^2 H^{2}\simeq\rho_{r}$.
This provides $(t-t_{\mathrm{os}})$ and, consequently the final $\rho_{r}$
as
\begin{equation}
t-t_{\mathrm{os}}\simeq\frac{160\pi}{3g_{*}\lambda^{2}m},\qquad\rho_{r}\simeq\frac{3g_{*}^{2}\lambda^{4}m^{2}\Mpl^{2}}{6400\pi^{2}}.
\end{equation}
The energy density of the produced particles is converted to the standard
expression for the energy density of a radiation gas as in $g_{*}\pi^{2}T^4_{\mathrm{rh}}/30$.
Therefore the reheating temperature can be estimated as
\begin{equation}
T_{\mathrm{rh}}\lesssim\lambda\left(\frac{9g_{*}}{640\pi^{4}}\right)^{1/4}\Mpl\left(\frac{m}{\Mpl}\right)^{1/2}.
\end{equation}
For the $f(R)$ gravity model $f(R)=R+R^{2}/(6m^{2})$, reheating
temperature takes instead the form
\begin{equation}
T_{\mathrm{rh}}\lesssim\left(\frac{g_{*}}{2560\pi^{4}}\right)^{1/4}\Mpl\left(\frac{m}{\Mpl}\right)^{3/2}.
\end{equation}
By comparing these two results, we find that the reheating temperature
of the 3-form field is larger than the one of the $f(R)$ gravity
model by $\Mpl/m\sim10^{5}$. Note that we estimated the value of 3-form mass, $m$, by using the power spectrum of the curvature perturbation found in \cite{Koivisto:2009fb} and using the observation data from seven-year WMAP \cite{WMAP7}. For the simple scalar field model with three-leg
interaction, the reheating temperature takes the form
\begin{equation}
T_{\mathrm{rh}}\lesssim g^{2}\sigma\left(\frac{9g_{*}}{40\pi^{4}}\right)^{1/4}\Big(\frac{\Mpl}{m}\Big)^{1/2}\sim\sigma.
\end{equation}
The result from this last calculation is comparable to the result calculated
by using the decay rate obtained from particle theory \cite{Abbott:1982hn,Albrecht:1982mp,Kofman:1994rk,Bouatta:2010bp}.


\subsection{Parametric resonance}

We investigate here the contribution of parametric resonances to the total energy density of produced matter particles in the 3-form inflation discussed here. The majority of the energy for the 3-form field, at the end of inflation,
is stored in the $k=0$ mode of the field. When the 3-form field oscillates
around the minimum of the potential, its energy undergoes coherent oscillations. This dynamics is the same as the one for a standard scalar
inflaton field.
In order to take into account the coherent nature of the 3-form field
at the end of inflation, we follow the standard procedure by investigating
the parametric resonance of the system \cite{Kofman:1994rk,Shtanov:1994ce,Kofman:1997yn}.
We will begin by introducing a new variable $y_{k}=a^{3/2}\phi_{k}$.
Therefore, Eq.~(\ref{eq:phik}) can be rewritten as
\begin{equation}
\ddot{y}_{k}+\left(\frac{k^{2}}{a^{2}}+m_{\phi}^{2}+\lambda Y-\frac{9}{4}H^{2}-\frac{3}{2}\dot{H}\right)y_{k}=0\,.\label{y_k1}
\end{equation}
In the limit $k^{2}\gg H^{2}\sim\dot{H}$, one can neglect the last
two terms of the above equation. Note that we still neglect the backreaction
of the 3-form quantum field in the following. Thus, we can still use
the solution of $Y$ as in Eq.~(\ref{Ysolution}). Substituting the
solution of $Y$ into Eq.~(\ref{y_k1}), we obtain
\begin{equation}
\ddot{y}_{k}+\left(\frac{k^{2}}{a^{2}}+m_{\phi}^{2}+\lambda\frac{\sqrt{8}\Mpl\sin[m(t-t_{\mathrm{os}})]}{\sqrt{3}(t-t_{\mathrm{os}})}\right)y_{k}=0\,.\label{y_k2}
\end{equation}
The parametric resonance occurs due to the oscillating term. In order
to see this behavior, we will introduce a new variable, $z$, defined
by $m(t-t_{\mathrm{os}})=2z-\pi/2$. Then Eq.~(\ref{y_k2}) is changed
to the Mathieu equation \cite{Mathieu} as follow
\begin{equation}
\frac{d^{2}y_{k}}{dz^{2}}+\Big(A_{k}-2q\cos(2z)\Big)y_{k}=0\,,\label{eq:Mathieu}
\end{equation}
where
\begin{equation}
A_{k}=\frac{4k^{2}}{a^{2}m^{2}}+\frac{4m_{\phi}^{2}}{m^{2}}\,,\qquad q=\frac{4\sqrt{8}\lambda\Mpl}{\sqrt{3}m^{2}(t-t_{\mathrm{os}})}\,.
\end{equation}
According to the Mathieu equation, there are instability bands in which
the perturbation $y_{k}$ grows exponentially with the growth index
$\mu_{k}=q/2$. These instability bands depend on the parameters $A_{k}$
and $q$. For a broad resonance, these parameters must satisfy the conditions
$A_{k}\simeq l^{2}$ and $q\gg1$ where $l^{2}=1,2,3,...$. For narrow
resonance, these parameter must satisfy the conditions $A_{k}\simeq l^{2}$
and $q\ll1$ where $l^{2}=1,2,3,...$. To guarantee enough efficiency
for the production of particles, the Mathieu equation's parameters
should satisfy the broad-resonance condition. However, in general,
the parameter $q$ decreases in time. Thus $q$ must take a large
enough initial value.

For the $f(R)$ gravity model, the parameters take the form \cite{DeFelice:2010aj}
\begin{equation}
A_{k}=\frac{4k^{2}}{a^{2}m^{2}}+\frac{4m_{\phi}^{2}}{m^{2}}\,,\qquad q=\frac{8\xi}{m(t-t_{\mathrm{os}})}\,.
\end{equation}
In order to get a broad resonance, $q\gg1$, the coupling constant
$\xi$ must take a large value. However, for the 3-form model, the
coupling constant does not need to be large, since $q\sim\lambda\Mpl/m$.

For a standard scalar inflaton field with four-legs interaction, the parameters take the
form \cite{Kofman:1994rk,Shtanov:1994ce,Kofman:1997yn}
\begin{equation}
A_{k}=\frac{k^{2}}{a^{2}m^{2}}+\frac{4m_{\phi}^{2}}{m^{2}}+2q\,,\qquad q=\frac{2g^{2}}{3m^{2}(t-t_{\mathrm{os}})^{2}}\,\frac{\Mpl^{2}}{m^{2}}.
\end{equation}
This model naturally gives large values for $q$, initially. However,
it decreases faster than the 3-form model, so that the broad resonance
tends to disappear more quickly.

Since the parameter $q$ is initially very large, the field passes through
many instability bands. This behavior leads to a stochastic change
in the growth index $\mu_{k}$. Therefore, in this situation, we need
to analyze the system as a stochastic resonance \cite{Kofman:1997yn}.
While the above discussion is based on the assumption that the backreaction
is negligible, it will be of interest to take into account the effect
of backreaction for more realistic model. The exponential growth of
the field also provides the non-adiabaticity in the change of the
frequency $\omega_{k}^{2}=k^{2}/a^{2}+m_{\phi}^{2}+\sqrt{8/3}\lambda\Mpl\,\sin[m(t-t_{\mathrm{os}})]/(t-t_{\mathrm{os}})$.
It should be noticed that, during the parametric resonance regime, the
produced particles are far away from equilibrium. The study of the thermalization
at the end of the parametric resonance regime is also of interest, and
we leave it for a future project.

\section{Conclusions}

The inflationary paradigm has joined cosmology with high energy particle
physics. In particular, high-energy theories may leave their footprint
in the data due to peculiar properties of the field which drives inflation.

In this paper we have investigated the properties of reheating due
to the presence of a 3-form field coupled with a matter field.
We have found that the existence of a coupled matter field does not change the conditions
that are necessary for ghost-free and Laplacian stabilities. However, to avoid a ghost field to appear,
the matter field has to be sub-dominant during inflation, but this condition is naturally fulfilled for quite a large variety of dynamics. The sub-dominance of a matter field is not required after inflation so that
reheating may occur without ghost and Laplacian instabilities, leading to a successful production
of relativistic particles at a scale
\begin{equation}
T_{\mathrm{rh}}\lesssim\lambda\left(\frac{9g_{*}}{640\pi^{4}}\right)^{1/4}\Mpl\left(\frac{m}{\Mpl}\right)^{1/2}.
\end{equation}
We have investigated also the production of particles due to parametric
resonances. In particular we have found that this process can be still
modeled by the Mathieu equation. Furthermore, compared to the standard
minimally-coupled scalar inflaton field scenario, reheating is more efficient,
as broad resonances typically survive longer, without the need
of choosing extremely large values for the coupling constant which models the interaction between the 3-form and the (relativistic) ordinary matter fields.
Finally, we have shown that indeed, the
production of particle at the end of inflation, due to the peculiar phenomenology of reheating,
gives us a possibility to distinguish experimentally such
a model from others alternative and viable models of inflation.

\begin{acknowledgments}
K.K.~is supported by Thailand Research Fund (TRF) through grant RSA5480009.
\end{acknowledgments}

\end{document}